# Stochastic modeling of hyposmotic lysis and characterization of different osmotic stability subgroups of human erythrocytes

Adriano Francisco Siqueira[1], Morun Bernardino Neto[1], Ana Lucia Gabas[1], Luciana Alves de Medeiros[3], Mario da Silva Garrote-Filho[3], Ubirajara Coutinho Filho[2], Nilson Penha-Silva[3]

[1]Engineering School of Lorena, University of São Paulo, Lorena, SP, Brazil
[2]Faculty of Chemical Engineering, Federal University of Uberlândia, Uberlândia, MG, Brazil
[3]Institute of Biotechnology, Federal University of Uberlândia, Uberlândia, MG, Brazil

Corresponding Author:
A. F. Siqueira, Email: adrianoeel@usp.br

## Abstract

This study proposes a novel stochastic model for the study of hyposmotic hemolysis. This model is capable of reproducing both the kinetics in the transient phase and the lysis equilibrium in the stationary phase, as well as the variability of the experimental measurements. The stationary distribution of this model can be approximated to a normal distribution, with mean and variance related to the salt concentration used in the erythrocyte osmotic fragility assay. The proposed model can generalize the classical Boltzmann sigmoidal model often used in adjusting the stationary experimental data distribution. A typical osmotic fragility curve is constructed from the absorbance of free hemoglobin as a function of the decrease in NaCl ($X$) concentration and allows the determination of $H50$, an osmotic fragility variable that represents the saline concentration capable of promoting 50% lysis, and $dX$, an osmotic stability variable that represents ¼ of the variation in salt concentration required to promote 100% lysis. Based on the stationary distribution of the proposed model it is possible to stratify a population into different groups of individuals with similar levels of cell stability. These groups are very suitable to study the factors associated with cell stability, such as gender, age and lipids, among others. The method presented here was applied to a sample of 71 individuals and several results were obtained. In a group of 25 female subjects, with $H50$ values between 0.42 and 0.47 g/dL NaCl, for example, the use of a quadratic model to study the dependence of the stability index $dX/H50$ with blood LDL-cholesterol levels, showed that the erythrocyte osmotic stability increases with increasing LDL-C to a maximum value close to 90 mg/dL and then decreases.

## Introduction

The erythrocyte is constantly exposed to several stressors, which may promote changes in its properties, such as deformability, which tends to decrease due to aggression promoted by reactive oxygen species (Cimen, 2008), for example.

Red cell deformability may also be affected by changes in membrane cholesterol content (Cooper, Arner, Wiley, & Shattil, 1975; Hui, Stewart, Carpenter, & Stewart, 1980; Lee, Kim, Park, Song, & Lee, 2004), promoted by the interaction of this cell with plasma lipoproteins (Hung, Berisha, Ritchey, Santore, & Smith, 2012; Quarfordt & Hilderman, 1970), in a process that will also affect other cellular properties, such as osmotic stability (Orbach, Zelig, Yedgar, & Barshtein, 2017).

In turn, the osmotic stability of erythrocytes is related to several factors, which include the parameters and indices provided by a complete blood count, such as total hemoglobin (Hb) concentration, mean corpuscular hemoglobin concentration (MCHC), mean corpuscular volume (MCV), red-cell distribution width (RDW) and erythrocyte count (RBC) (M. V. de Freitas et al., 2014; Paraiso et al., 2017).

The relationship of erythrocyte osmotic stability with RDW is noteworthy, since RDW elevation has been able to predict the aggravation of several diseases (Patel et al., 2010; Perlstein, Weuve, Pfeffer, & Beckman, 2009; Zurauskaite et al., 2018). A broader understanding of the factors that affect the erythrocyte osmotic stability should certainly shed more light on the pathological mechanisms associated with the predictive ability of RDW.

In addition to osmotic stability, the kinetics of erythrocyte hyposmotic lysis, studied by Cunha and colleagues with the help of the classic Michaelis-Menten kinetic model, has also been shown to have important relationships with age, hematimetric parameters and lipidemia (Cunha et al., 2014). An empirical model composed of two exponential functions was also proposed by Gornicki to study hemolysis kinetics (Gornicki, 2008). However, none of these works presents a theoretical mathematical modeling for experimental variations of hemolysis rate.

The aim of this paper was to present an unpublished stochastic model for the study of erythrocyte hyposmotic lysis. This model is capable of reproducing the lysis kinetics in the transient phase and the lysis equilibrium in the stationary phase, as well as the variability of experimental measurements. This enabled the development of a new method to group individuals with similar values of osmotic fragility. One among many possible applications of this method, the investigation of the influence of hematimetric and biochemical parameters on the erythrocyte stability was explored in this study, in order to evaluate the potential of the model.

This is very relevant, since the osmotic stability of erythrocytes is an important clinical tool that has been applied not only in the diagnosis of erythrocytopathies, but also in the study of numerous diseases, such as diabetes mellitus (Rodrigues et al., 2018), preeclampsia (Aires Rodrigues de Freitas et al., 2018; M. A. R. de Freitas et al., 2019) and malaria (Mascarenhas Netto Rde et al., 2014), in processes such as aging (M. V. de Freitas et al., 2014; Penha-Silva et al., 2007) and clinical interventions such as physical exercise (Paraiso et al., 2017) and bariatric surgery (de Arvelos et al., 2013), among other situations.

## Material and Methods

### Collection of blood samples

The blood samples used in this study were residual aliquots of blood from subjects who cannot be identified or contacted even if the results obtained could benefit them. Samples were collected by venipuncture directly into evacuated tubes containing 1g/dL $K_3$EDTA as an anticoagulant (Vacutainer, Becton Dickinson, Juiz de Fora, MG, Brazil) after 8-12 hours fasting, in a clinical laboratory (LABORMED) of Uberlândia, MG, Brazil, and stored at 0 to 4 °C for a maximum of 24 hours before testing.

### Reagents

All reagents used in the experiments had ACS grade of purity. The purity of NaCl (Labsynth, Diadema, SP, Brazil) used here was 99.5%, which was duly corrected in the preparation of its solutions.

**Equipment**

All weighings were done using a digital precision scale (AND, model 870, Japan) and volume measurements were made using automatic pipettes (Labsystems, Finnpipette Digital model, Helsinki, Finland). Incubations were performed in a thermostatic bath (Marconi, model MA 184, Piracicaba, SP, Brazil). Centrifugations were performed in a temperature controlled centrifuge (Hitachi Koki, model CF15RXII, Hitachinaka, Japan) and absorbance readings were taken on a UV-VIS spectrophotometer (Shimadzu, model UV1650TC, Japan).

Blood counts were obtained on an automated system (Sysmex K4500, Sysmex Corporation, Mundelein, IL, USA). Total cholesterol (t-C), HDL-cholesterol (HDL-C), LDL-cholesterol (LDL-C), VLDL-cholesterol (VLDL-C), triglycerides (TGC), and glucose (GLU) determinations were performed by an automated analyzer (Hitachi 917, Roche Diagnostics, Indianapolis, IN, USA).

**Experimental determination of erythrocyte osmotic stability**

A duplicate series of test tubes containing 1 mL of 0.1-1.0 g/dL NaCl solution was previously incubated at 37 °C in a thermostated water bath for 10 minutes. After adding 10 μL of whole blood to all tubes, they were gently shaken and incubated again at 37 °C for 30 minutes. The samples were then centrifuged at 1600 x $g$ for 10 minutes for supernatant separation and absorbance reading at 540 nm.

The osmotic fragility curve was constructed for the dependence of the absorbance at 540 nm (A) as a function of the NaCl (X) concentration, using a sigmoidal regression routine based on the Boltzmann equation,

$$A = \frac{A_{max} - A_{min}}{1 + e^{(X - H_{50})/dX}} + A_{min} \qquad (1),$$

in order to determine the values of the parameters *Amax* and *Amin*, which represent the absorbance values at 540 nm in the maximum and minimum sigmoid plateaus; *H50*, which represents the NaCl concentration capable of causing 50% hemolysis; and *dX*, which represents ¼ of the variation in NaCl concentration responsible for the lysis of the whole erythrocyte population used in the assay (Bernardino Neto et al., 2013; Penha-Silva et al., 2007).

**Mathematical Modeling**

In assays performed to determine the erythrocyte osmotic stability, aliquots of blood are added at different NaCl concentrations and the suspensions obtained are incubated for 30 minutes, a time interval which is sufficient for the occurrence of different lysis rates depending on the medium osmolarity. After this time, each mixture contains a specific amount of free hemoglobin in solution obtained from hemolysis under the osmolarity condition present therein. After centrifugation and separation of the supernatant, the measured absorbance expresses the free hemoglobin concentration, which does not strictly reflect the amount of cells lysed, since a red blood cell population shows heterogeneity in hemoglobin concentration and deformability. This is due to several factors such as the existence of differences in the lifetime of these cells, because as red cells age, younger cells, which are larger and more deformable, lose hemoglobin and become less deformable (Clark, 1988; Franco et al., 2013; Malka,

Delgado, Manalis, & Higgins, 2014). In addition, the heterogeneity of the erythrocyte population is greatly influenced by the conditions of supply of nutritional factors such as folate, cobalamin, pyridoxine, iron and proteins (Koury & Ponka, 2004). All of this must mean that the hemolysis rate, given by $r = n/N$, where $n$ is the number of cells in lysis and $N$ is the initial number of cells in the suspension, is difficult to measure at any given moment because it is under the influence of various factors that are difficult to control or even unknown, in addition to experimental variability. Therefore, $r$ was considered a random variable in the discrete probabilistic model for the increase in hemolysis rate proposed here, ie,

$$\Delta r = \begin{cases} & Probability \\ -\gamma & k2*n*\Delta t \\ +\gamma & k1*(N-n)*\Delta t \\ 0 & 1 - \big(k2*n + k1*(N-n)\big)*\Delta t \end{cases} \quad (2),$$

where $\gamma$ is a proportionality constant, $n$ is the number of cells lysing at a given time, $N$ is the number of intact cells at the initial time of mixing, $k1$ and $k2$ are constants that depend on salt concentration, and $\Delta t$ is a small time interval. This model assumes that the hemolysis rate depends on the number of intact cells and the saline concentration in each solution used in the assay.

In other words, the described model (Equation 2) assumes that the hemolysis rate may increase by $+\gamma$, with probability proportional to the number of intact cells, or decrease by $-\gamma$, with probability proportional to the number of cells that have been lysed. Thus, the greater the number of intact cells, the greater the chance hemolysis rate increase. On the other hand, the greater the number of lysed cells, the greater the chance of hemolysis rate decrease. Thus, the rate of hemolysis may increase or decrease at any moment, like a car always going in a certain direction, but changing speed, sometimes faster, sometimes slower. Both probabilities also depend on the saline conditions of the experiment, indicated by the constants $k1$ and $k2$. Thus, for each time increment, the hemolysis rate may increase or decrease according to the probabilities described in the model (Model 1).

This discrete stochastic model (Equation 2) follows the recommendations of chapter 5 of Allen (Allen, 2007) and can be approximated by a stochastic differential equation that has approximately the same probability distribution:

$$dr = \gamma N(k1(1-r) - k2r)dt + \gamma\sqrt{N}(\sqrt{k1(1-r) + k2r})dB \quad (3),$$

where $B$ is a replica of the Brownian movement.

Considering $\gamma = N^{-1}$ and $\varepsilon = 1/\sqrt{N}$, Equation 3 becomes:

$$dr = (k1(1-r) - k2r)dt + \varepsilon(\sqrt{k1(1-r) + k2r})dB \quad (4).$$

Further discussion of this approach is given in Appendix A.

The first component of this equation is a measure the tendency, or drift, of rate over time, while the second component gives, of the variable $r$. For $r \in [0\ 1]$, using the technique described by Grasman and van Herwaarden (Grasman & van Herwaarden, 2010), it is possible to prove that this model has a nearly normal stationary distribution, characterized by mean and variance shown in equations:

$$E(r) = \frac{k1}{k1+k2} \qquad (5)$$

and

$$dp(r) = \frac{\varepsilon\sqrt{k1k2}}{k1+k2} \qquad (6),$$

respectively.

This means that the hemolysis rate, given by $r$, will fluctuate over a long period of time around the average rate (Equation 5).

With the model described by Equation 4, the average signal associated with the measurement of absorbance at 540 nm ($\bar{A}$) can be studied in a simplified manner. For this we must consider that the red blood cell can generate two types of signals, the *signal D*, when it undergoes lysis, and *signal L*, when he does not suffer lysis (Figure 1).

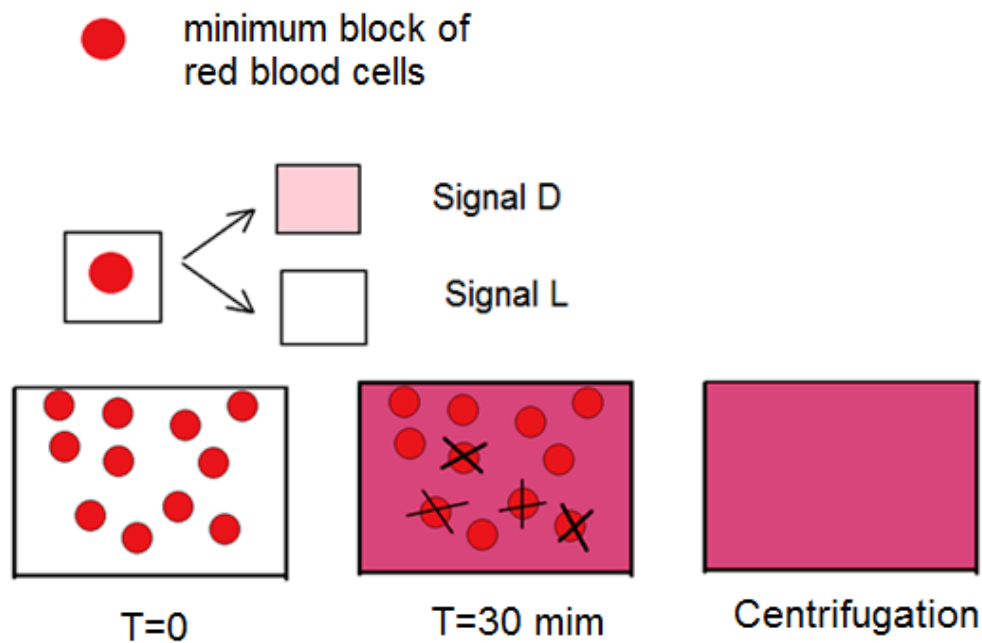

**Fig. 1**. Schematic representation of the mean signal obtained in the osmotic stability assay after centrifugation of the cell suspension.

Thus, the average signal can be approximated by the sum of *signal D* of a lysed cell block multiplied by the number of lysed cells, with the *signal L* of a non-lysed cell block multiplied by the number of non-lysed cells, all divided by $N$.

Thinking in terms of the detection limit of the apparatus, $N$ should be interpreted as a minimum block of cells that, when lysed, generate a detectable response by the equipment. Thus, a test tube can be thought of as being grouped by several of these minimal erythrocyte blocks. Thus, $N$ will depend on the sensitivity of the equipment used by the analysis.

Although *signal D* and *signal L* are presumably constant, regardless of the individuals in a particular group, they may vary between different groups. Signal differences due to differences in the amount of hemoglobin in each erythrocyte, among other factors, can be modeled by variations in hemolysis rate, given by $r$, through Equation 6.

All these assumptions were considered in an equation that allows the calculation of the average signal:

$$\bar{A} = \frac{E(Signal)}{N} = \frac{Signal\,D*N*E(r)+Signal\,L*N*(1-E(r))}{N} = Signal\,L + (Signal\,D - Signal\,L) * \frac{k1}{k1+k2} \quad (7).$$

The standard deviation of the average signal (SD$_{\bar{A}}$) can also be calculated using the following equation:

$$SD_{\bar{A}} = (Signal\,D - Signal\,L) * \varepsilon * \frac{\sqrt{k1k2}}{k1+k2} \quad (8).$$

Although the number of cells, indicated by *N*, are important to the model described by Equations 1, 2 and 3, the final expression used for calculating the expected average value of absorbance (Equation 7) does not depend explicitly on *N*, although *signal D* and *signal L* depend on the amount of hemoglobin, which is related to the number of cells or initial cell blocks used in the osmotic fragility assay and also to the spectrophotometer's detection capability.

*Choice of k1 and k2*

The choice of constants *k1* and *k2*, which depend only on the salt concentration of each assay, can be made by different criteria, as was done in the following models:

Model I: $k1{=}1$ and $k2 = exp(\frac{X-H50}{dX})$,

where *X* is the salt concentration, *H50* is the salt concentration that causes a hemolysis rate (*r*) of 50% and *dX* is a model adjustment constant;

Model II: $k1{=}a$ and $k2 = exp(\frac{X-H50}{dX}){+}$b,

where the positive constants *a* and *b* satisfy the ratio *a - b = 1*, so that the expected value (*E*) of hemolysis rate (*r*) given by Equation 5 is 50% when *X = H50*; and

Model III: $k1{=}a$ and $k2 = \frac{100H5}{1+\left(\frac{H5}{X}\right)^n} + b$,

where *a*, *b* and *n* are positive constants, *H5* is the salt concentration responsible for *r* = 5%, and *a* e *b* satisfy the equation 0.95*a* - 0.05*b* = 2.5 *H5* so that the expected value (*E*) of hemolysis rate (*r*), given by Equation 5, be 5% when *X = H5*.

Model I was built based on the classical model of Boltzmann (Dubois, Ouanounou, & Rouzaire-Dubois, 2009). Indeed, considering *A$_{max}$ = Signal D* and *Amin = Signal L,* and substituting the values of *k1* and *k2* of this model in Equation 5, this equation is exactly equal to the Boltzmann equation:

$$\bar{A} = Signal\,L + \frac{Signal\,D - Signal\,L}{e^{(X-H50)/dX}} \quad (9).$$

Model 2 is an adaptation of Model 1 and Model 3 is an adaptation of the Hill model (Yasuhara & Kuroda, 2015).

*Confidence Intervals*

Since equilibrium distribution of the hemolysis rate (*r*) given by Equation 4 approximates a normal distribution, with mean given by Equation 5 and standard deviation given by Equation 6, it is possible to construct a confidence interval (*CI*) for *r*, with c% confidence, where typically *c* = 95, based on the following equation:

$$CI(c\%) \approx \frac{k1}{k1+k2} \pm \frac{\varepsilon z_c \sqrt{k1k2}}{k1+k2} \qquad (10).$$

Equation 10 can also provide different estimates of *H50,* when confidence interval takes 50% the hemolysis rate, as illustrated in Figure 2.

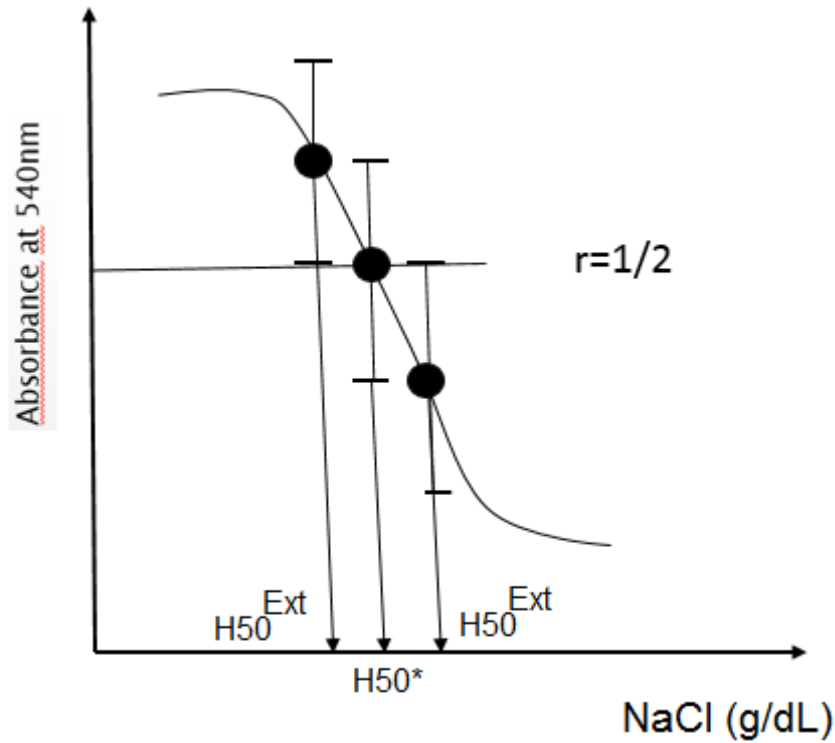

**Fig. 2**. Schematic representation of the confidence intervals given by Equations 11 and 12.

The salt concentration whose *CI(c%)* of hemolysis rate (*r*) equals ½ represents an estimate of the true *H50* of the studied population. This *H50* is identified in Figure 2 as *H50\**.

*Groups Analysis*

Considering only *H50* as a measure of cellular fragility, Equation 8, which allows the calculation of the hemolysis rate confidence interval, can be used to group individuals based on the proximity of *H50* values.

After determining the *CI* for a given individual's *H50* value, the $H50^{Ext}$ values shown in Figure 2 can be estimated by the equations:

$$H50^{Ext}\ max = H50^* + 2z_c dX\varepsilon \qquad (11)$$

and

$$H50^{Ext} \, min = H50^* - 2z_c dX \varepsilon \qquad (12),$$

which are valid for Models I, II and III, since data dispersion of absorbance in average NaCl concentration region, is similar in these three situations.

Thus, all *H50* values within the range limited by Equations 11 and 12 can be considered similar.

The *H50\** and *dX* values present in Equations 11 and 12 are those obtained in the subject's cell fragility assay.

The only parameter with unknown value in those equations is ε, derived from the stochastic model proposed in Equation 4. This parameter is related to the variability of cell fragility measurements, due to gender and age differences, among other factors. But we can point out three alternatives for estimating the value of ε.

The first alternative involves the kinetics of hyposmotic erythrocyte lysis (Cunha et al., 2014) at various time periods within the 30 minutes used in the cell fragility assay of the present study.

Assuming that a particular set of individuals forms a group with common characteristics (such as gender, age, and pathological condition), a second alternative involves only the cell fragility assays to estimate the values of ε and the parameters of Model I, II or III. This can be done by the best fit of the expected values and standard deviation of each model with the mean values and standard deviation of the experimental data.

A third alternative is to study, from the variation of ε, the possible groups existing in a set of individuals with cell fragility determined by Equation 4. In the hypothesis of a group of individuals presenting a chosen value of ε, it is possible to identify the individuals that are part of this group from the similarity of the values of their cell fragility parameters.

Once a group of individuals has been characterized, it is possible to study, for example, the relationships between cell stability parameters, such as *dX* and *H50*, and even cell stability indices, such as *dX/H50*, with blood hematimetric and biochemical parameters, among others.

*Lysis Time*

In a study by Górnicki (Gornicki, 2008), the dependence of transmittance, *T*, as a function of time, *t*, is used to describe the average behavior of erythrocyte lysis kinetics, according to the empirical model given by the equation:

$$T(t) = A - B exp\left(\frac{-t}{t_B}\right) - C exp(\frac{-t}{t_C}) \qquad (13),$$

where *A* represents lysis of the total cell population, *B* is lysis of the most rapidly lysed cell fraction, and *C* is lysis of the most slowly lysed cell fraction, such that $A = B + C$; and $t_B$ and $t_C$ represent the fast and slow lysis times, respectively.

In terms of comparison, when erythrocyte geometry does not change, the author suggests that $t_B$ and $t_C$ may be related to fluidity and other hemolytic properties of the cell membrane, such as permeability to monovalent cations, such that the cell membrane is both less fluid, ie, more rigid, the longer it takes to reach steady state.

The average behavior of the hemolysis rate *(r)* given by Equation 4 over time *(t)*, ie r(t) , can be given by the equation:

$$E\big(r(t)\big) = \frac{k1}{K1+k2}(1 - \exp(-(k1+k2)t)) \tag{14}.$$

The Taylor series expansion of the exponential terms of Equation 13 for small time intervals allows Equation 15 to approximate the following equation:

$$T(t) \approx A\left(1 - exp\left(\frac{-\left(\frac{Bt_C + Ct_B}{A}\right)t}{t_B t_C}\right)\right) \tag{15}.$$

Since the transmittance measurements of the Górnicki experimental assays were made without centrifugation and, under this situation, *T(t)* is proportional to *r(t)*, it is then possible to compare Equation 14 with Equation 15. Thus, it is clear that the constant A in (Equation 13) is the same as that given by the average value of the steady state (Equation 14). Furthermore, an analysis of Equation 15, the term $t_B t_C$ suggests dividing the constants *k1* and *k2* of the Equation 2 model by a new constant, called *Lt,* related to the lysis time. This operation allows a better control of the dynamic stage of the lysis given by Equation 4 and, consequently, by the other models derived from it, without changing the average value and standard deviation of *r(t)* at the steady state, Equations (5) and (6). These new *k1* and *k2* constants, now called $k1^N$ and $k2^N$, are related by approximation of Equations 13 and 14, according to the following equation:

$$k1^N + k2^N = \frac{k1+k2}{Lt} = \frac{B}{A}\left(\frac{1}{t_B}\right) + \frac{C}{A}\left(\frac{1}{t_C}\right) \tag{16}.$$

The *Lt* parameter, which may be called overall lysis time to differentiate from $t_B$ and $t_C$, influences the dynamic behavior of lysis and may be related to some characteristic of an individual and/or a subgroup of individuals that are part of a larger group with the same steady state hemolysis rate. This new constant can be measured in dynamic experiments. This topic has been added to show that the model given by Equation 1 and subsequent equations can reasonably represent the dynamic mean behavior of cell lysis by incorporating a new constant related to lysis time. However, the discussion of the results of this study involving steady state will be considering the use of *k1* and *k2* instead of $k1^N$ and $k2^N$.

## Results and Discussion

### Sample Characteristics

The study population consisted of 45 female and 26 male subjects. The baseline characteristics of the total and female study population can be seen in **Table 1** and **Table 2**, respectively.

### Kinetic Behavior

Figure 3 shows some numerical simulations of the stochastic model of Equation 4, at different saline concentrations, using Model II values of *k1* e *k2*.

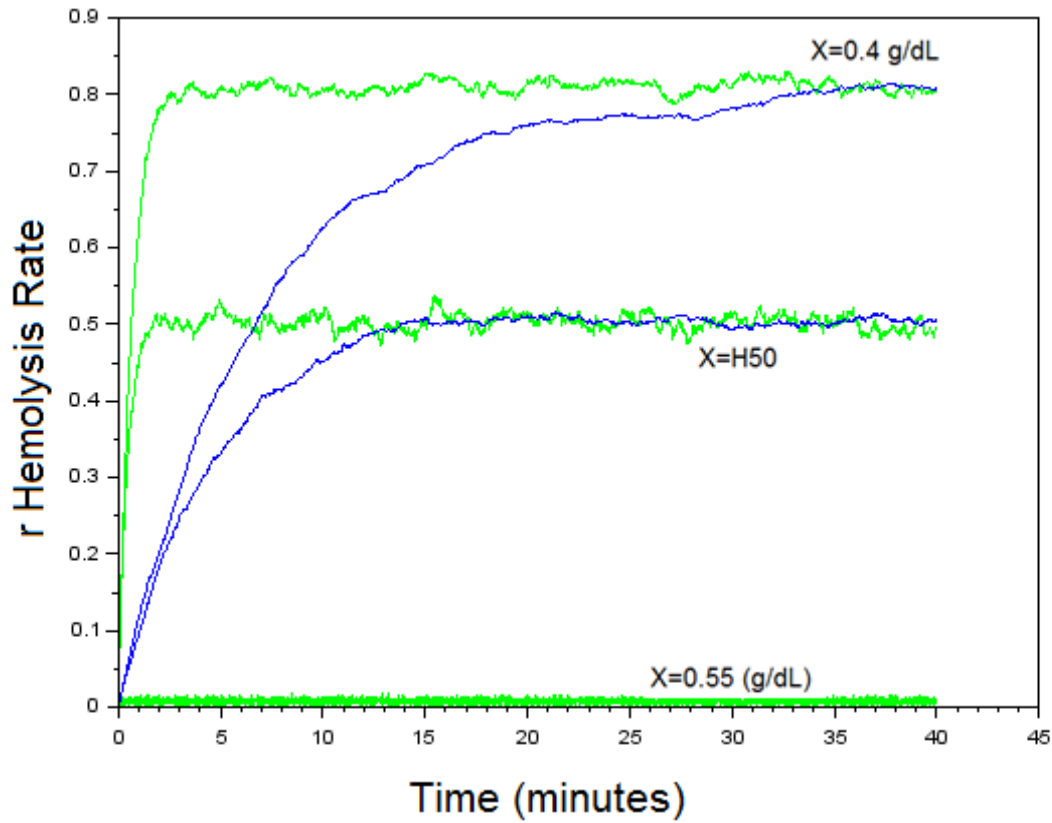

**Fig. 3**. Numerical simulation of the stochastic model given by Equation 3 for different salt concentrations using *H50* = 0.45, *dS* = 0.02, *a* = 1.2, *b* = 0.2 and *ε* = 0.02 in all situations and Lt values of 1 (green line) and 10 (blue line).

These simulations satisfactorily represent the temporal variation of the experimental hemolysis rate shown in Figure 1 of the study by Cunha and colleagues (Cunha et al., 2014), showing that the model considered in Equations 1-3 is able to adequately represent the kinetic behavior of the hemolysis rate. Furthermore, Figure 4 also shows the kinetic behavior described for hemolysis reported by Gornicki (Gornicki, 2008), showing that it is possible to control the speed with which the model reaches steady state using different values for the overall time of hemolysis (*Lt*).

**Groups Analysis**

Initially, the hypothesis that the group of 71 individuals analyzed constitutes a homogeneous population regarding cellular fragility was considered, only to show that the model can follow the variability of the measurements, even in a heterogeneous group regarding gender, age and other factors. The parameters of Models I, II and III were then identified as described in the Material and Methods section.

Figure 4 shows the experimentally obtained values for the mean (Figure 4A) and standard deviation (Figure 4B) of those 71 subjects, compared to the mean and standard deviation values of the absorbance signals calculated using Model II, shown as solid blue lines in both figures.

A)

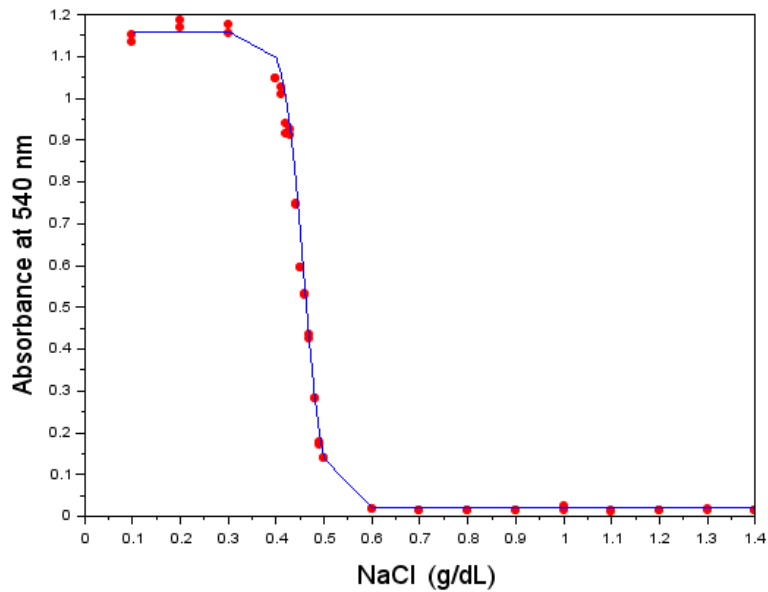

B)

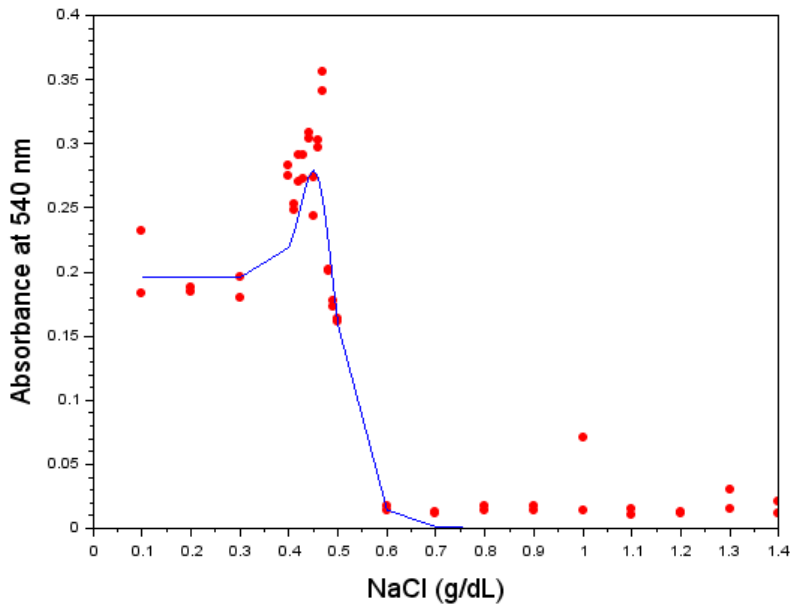

**Fig. 4**. Mean (**A**) and standard deviation (**B**) values of the absorbance signals that were experimentally measured in the cell fragility assays (red balls) compared to the mean and standard deviation values determined based on Model II (solid blue line).

The experimental data adjustments to Models I, II and III are shown in Figure 5.

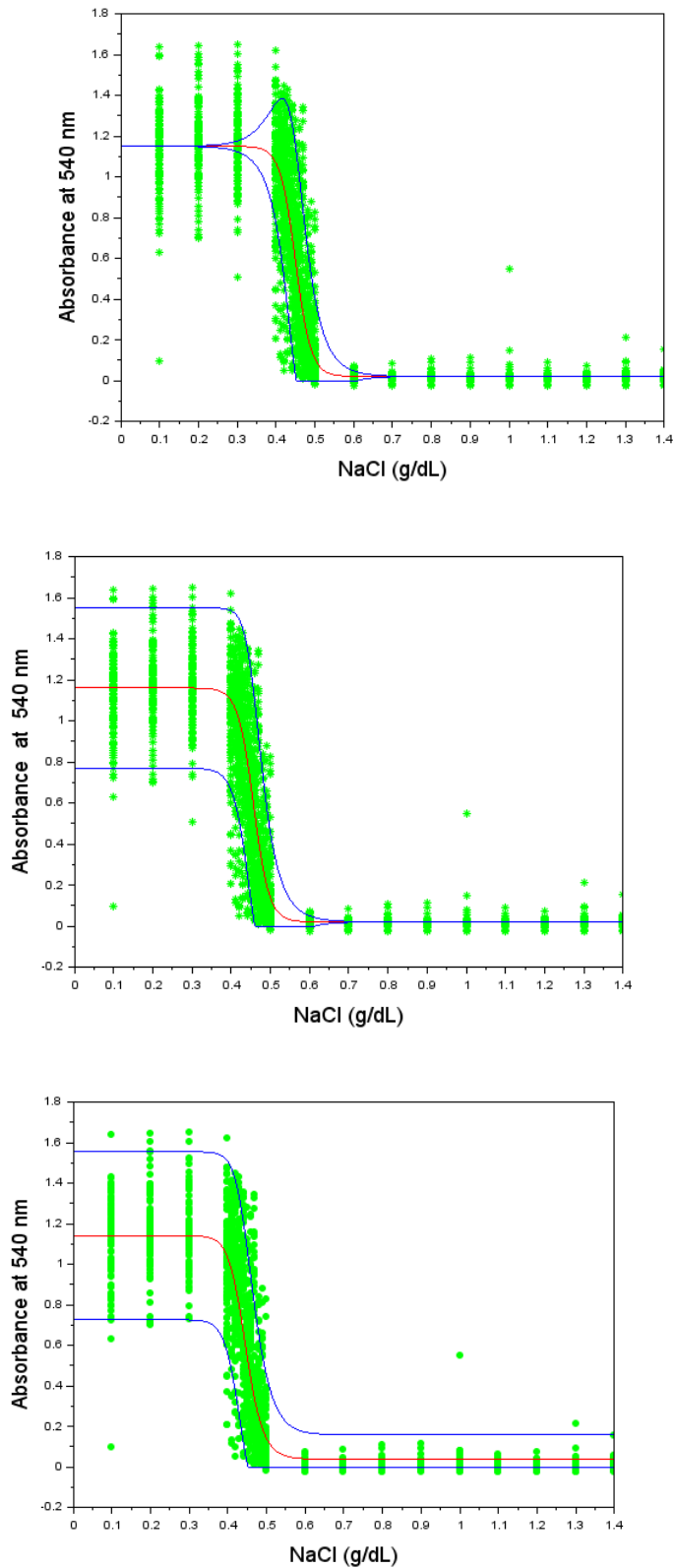

**Fig. 5**. Fitting of the proposed models: Model I: *H50* = 0.45, *dS* = 0.02, *ε* = 0.5, *Signal L* = 0.02 and *Signal D* = 1.15 (**A**); Model II: *H50* = 0.45, *dS* = 0.02, *a* = 1.2, *b* = 0.2, *ε* = 0.42, *Signal L* = 0.02 and *Signal D* = 1.35 (**B**); and Model III: *H5* = 0.55, *a* = 1.46, *b* = 0.24, *n* = 20, *ε* = 0.4,

*Signal L* = 0.04 and *Signal D* = 1.4 (**C**). The solid red line represents the mean value and the blue lines delimit the confidence interval (95%) for the signal.

All three models can represent the variability and mean value of the data in the *H50* region. However, Model I cannot represent data variability in the low and high NaCl concentration regions. Model II can represent the variability of the data in the region of low NaCl concentration, but also does not have good representation in the region of high concentration of NaCl. Finally, Model III can represent data variability and mean value in the three regions, that is, in the low, medium and high salt concentration regions. Barros and colleagues called these three regions of salt concentration phases in a study on the influence of different concentrations of essential oils on cell fragility in each of these phases (Barros et al., 2016). Certainly the characterization of individuals regarding these phases may contribute to the understanding of the influence of different factors on their hematimetric characteristics. In addition, the use of Confidence Intervals generated by Models I, II and/or III may help in planning the evaluation of osmotic fragility of a new individual from the phase that best describes his/her osmotic fragility curve. But above all, these results suggest the existence of heterogeneity in osmotic fragility of erythrocytes.

Considering now the hypothesis of heterogeneity in the osmotic fragility of the sample of 71 individuals studied here, given by the presence of subgroups with different levels of cellular fragility, the grouping technique proposed in the Material and Methods section was applied by the variation of the values of $\varepsilon$ between 0.05 and 0.5. Some results of this analysis were shown in Figure 6.

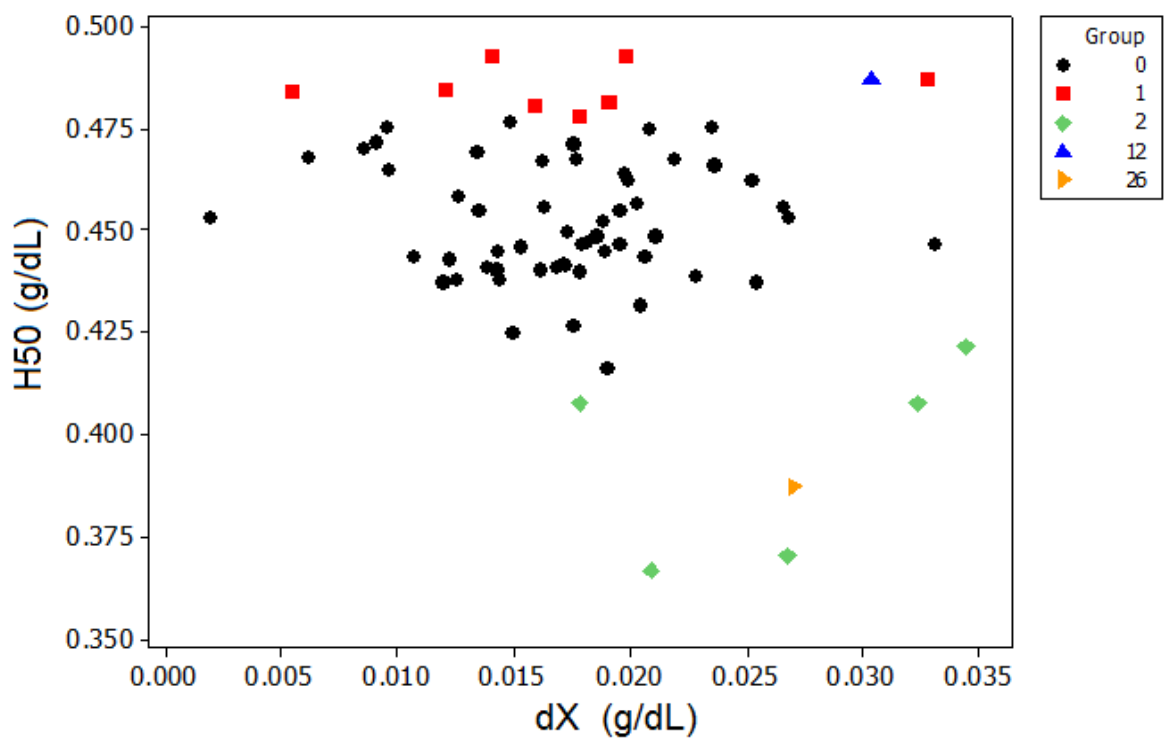

**Fig. 6**. Subjects whose *H50* values are similar to those of individual # 12 (group 1) and individual # 26 (group 2) compared to subjects whose values of *H50* are not similar to those two individuals (group 0).

This figure highlights individuals 12 and 26, who were chosen because they represent groups with higher and lower erythrocyte osmotic fragility, respectively. Thus, there would be a group of individuals similar to individual 12, with a value of ε = 0.1, and a group of individuals similar to individual 26, with ε = 0.25.

The osmotic fragility curves of the groups of individuals with higher and lower cell fragility has been presented in Figure 7, along with the average line and the lines of the 95% confidence interval which was determined based on Model I. The separation of the two groups in the intermediate salt concentration region is quite evident.

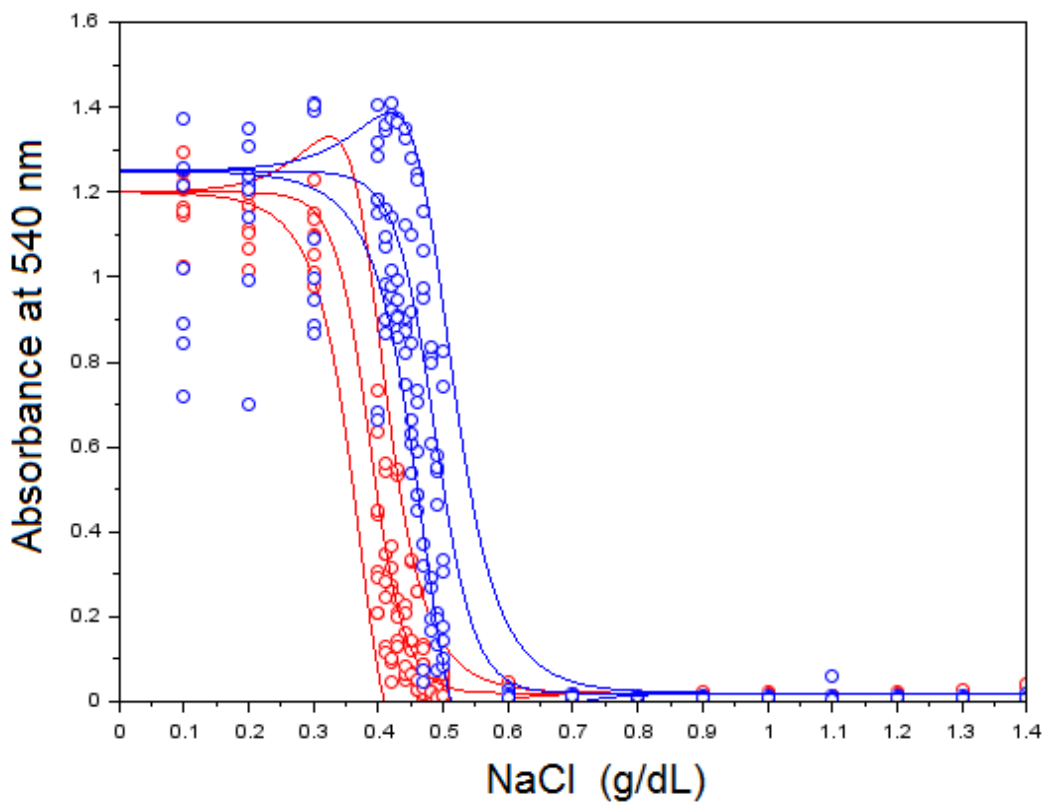

**Fig. 7**. Application of Model I (mean and 95% confidence interval) to the subgroups of major (blue circles) and minor (red circles) osmotic cell fragility of Figure 7.

Finally, to highlight the potential of the technique of grouping individuals by similarity in cell stability from variation in the value of the parameter ε, some relationships found by linear regression analysis are used. Values of ε between 0.05 and 0.5 were chosen for this purpose.

The first significant relation found is given by the following equation:

$$\frac{dX}{H50} = a_0 + a_1\,LDL + a_2 MCV * RDW \tag{17},$$

where *LDL*, *MCV* and *RDW* represent, respectively, the blood levels of LDL-cholesterol and the values of mean corpuscular volume and red-cell distribution width.

A group of 20 individuals with *H50* values between 0.4005 and 0.4424 was found for ε = 0.15, but the largest group found occurred with ε = 0.45 and *H50* values between 0.2959 and 0.4443. This group comprised 26 individuals of the sample and the linear regression analysis described by Equation 18 for this group, given by the following equation:

$$\frac{dX}{H50} = 0.136021 + 0.000202624\,LDL - 0.00009.62565\,MCV * RDW \qquad (18),$$

showed significance (*p* < 0.05) for all model constants. Several smaller subgroups were found in this range of *H50* and the regression analysis by Equation 17 always preserved the signals found for the values of *a1* and *a2* of Equation 18. But adjustment by Equation 17 was not statistically significant for both independent variables in any of the groups of individuals with *H50* values above 0.4443.

A clinical reading the results expressed by Equation 18 shows that for the 26 individuals with *H50* in the range between 0.2959 and 0.4443, the osmotic stability of erythrocytes increases with an increase in *LDL* cholesterol and decrease in the standard deviation of *MCV*, since the product of *MCV* and *RDW* results in the standard deviation value of *MCV*. Moreover, since it is unlikely that there is an indefinite enhance in the osmotic stability of erythrocytes with increased LDL-C levels, this shall mean that Equation 18 provides evidence that such increase in the standard deviation of *MCV* is restricting the enhance in stability of erythrocytes with the increase in LDL-C levels. Since this enhance in MCV standard deviation means increased RDW, and LDL-C levels somehow reflect the membrane cholesterol content, it is possible that Equation 18 reflects the behavior of osmotic stability in response to increased cholesterol content in the erythrocyte membrane. Indeed, elevation in *RDW* values was associated with elevation in erythrocyte membrane cholesterol content (Tziakas et al., 2012). This is immensely relevant as it helps support the idea that changes in erythrocyte membrane cholesterol content is directly associated with the pathophysiology of atherosclerosis and its complications (da Silva Garrote-Filho, Bernardino-Neto, & Penha-Silva, 2017).

Figure 8 highlights the relationship between *H50* and *dX* for group (group 1) where Equation 17 was significant in relation to *LDL* and *MCV∗RDW* compared to the other individuals in the sample (group 0).

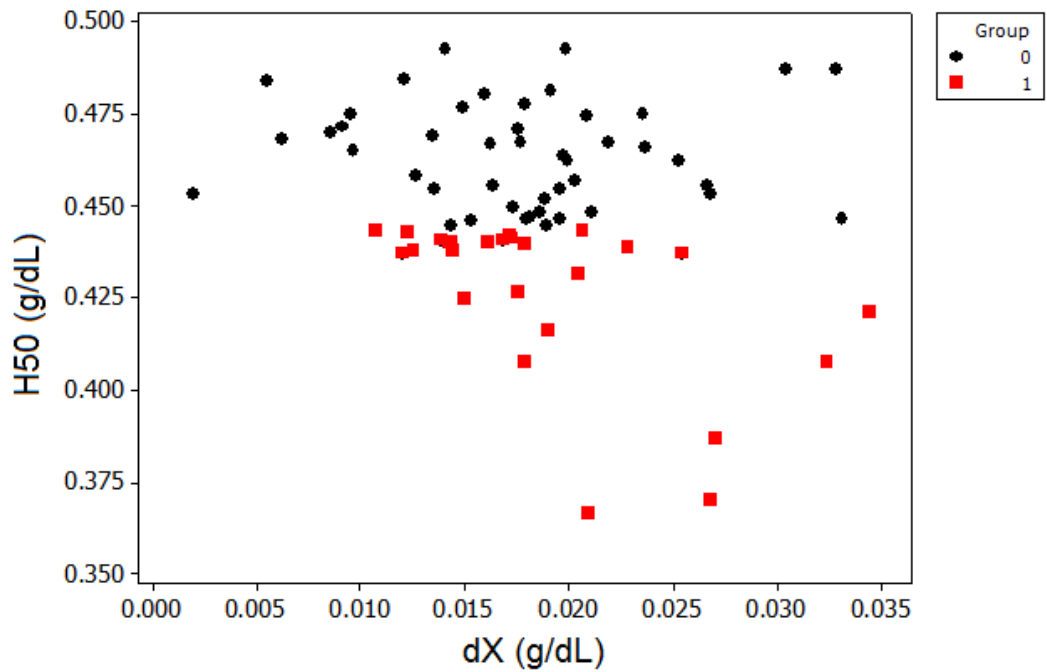

**Fig. 8**. Group of individuals for whom the model given by Equation 17 is significant (group 1) compared to other individuals (group 0).

The finding that LDL-C levels exert dualistic influence on erythrocyte stability deserves further analysis and therefore the sample of 71 subjects in this study was investigated for the existence of any group of individuals for which cell stability could be explained by a quadratic relationship with the *LDL* variable. The stratification of the sample by gender (45 females and 26 males) was able to identify in the group of females several subgroups in which the quadratic relationship between cell stability and *LDL* is significant among individuals with *H50* values between 0.42 and 0.47. The largest of these subgroups presents all female individuals (n = 25) found for values of $\varepsilon = 0.25$. In all these subgroups the quadratic term of *LDL* has negative sign and significant correlation with *dX/H50*. The quadratic relationship found ($p < 0.01$) is given by the equation:

$$\frac{dX}{H50} = -0.03143 + 0.001744\ LDL - 0.000010\ LDL^2 \qquad (19),$$

and the normality analysis of its residues by the Anderson Darling test does not reject the normality hypothesis ($p = 0.52$). Figure 9 shows this quadratic relationship between *dX/H50* erythrocyte stability index and LDL-C levels.

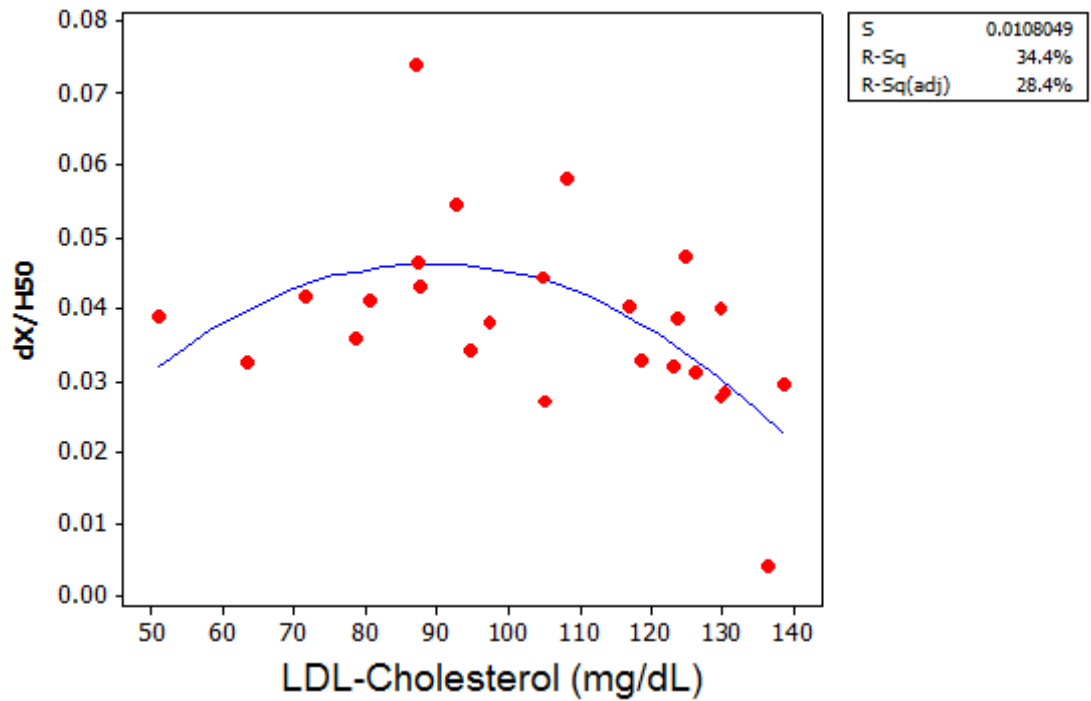

**Fig. 9**. Adjustment of the dependence of the osmotic stability index, *dX/H50*, with blood levels of LDL-cholesterol (Equation 17) for the female group identified by the model given by Equation 3.

Figure 10 highlights the relationship of *H50* values of these individuals as a function of their respective *dX* values compared to the other female subjects. This figure shows that Equation 19 is significant only for the female group with *H50* values in a central range (0.42-0.47) of the variability spectrum of this variable in the female population of the sample considered in this study.

**A)**

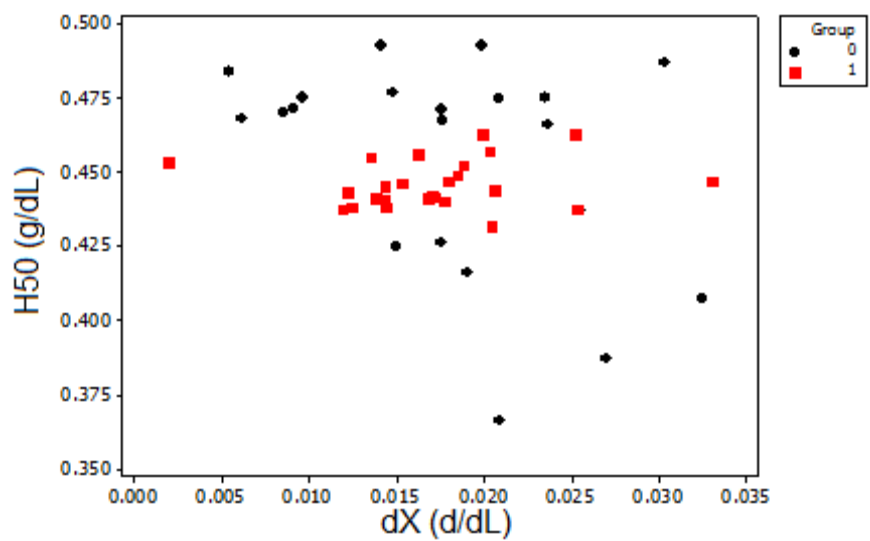

B)

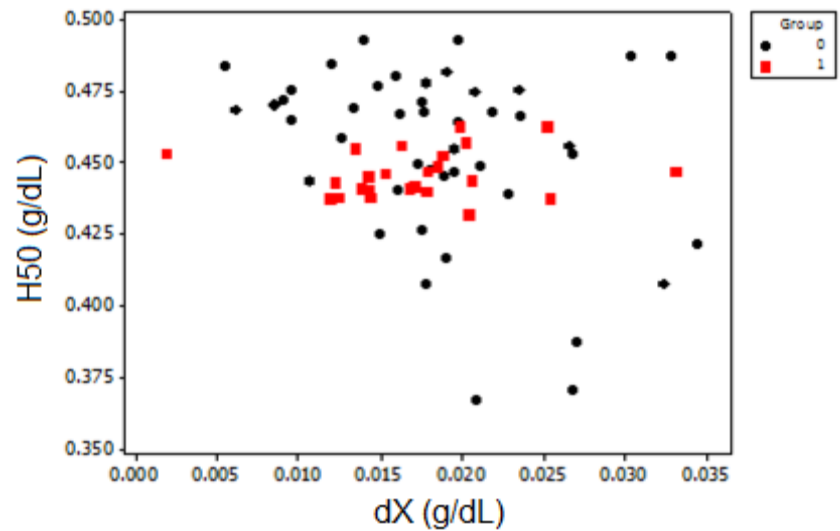

**Fig. 10.** Female subjects for whom Equation 19 model is significant (group 1) compared to the other female subjects (group 0) (**A**) and all subjects of both sexes (**B**) for whom that model is not significant.

The existence of a dualistic influence of LDL-C levels on erythrocyte stability is a very relevant issue and may help to understand the source of the antagonistic influence that blood cholesterol levels exert on blood cell count (Fessler, Rose, Zhang, Jaramillo, & Zeldin, 2013).

**Conclusions**

The stochastic model proposed here can reproduce the mean profile and data variability of the erythrocyte osmotic fragility assay of a group of individuals. This unprecedented technique allows to group individuals with similar values of cell fragility and from these groups to study the possible factors associated with erythrocyte stability, such as gender and erythrogram and lipidogram variables, for example. Moreover, the representation of lysis kinetics with a parameter called lysis time makes it possible to control the speed at which the rate of hemolysis reaches steady state.

**Appendix A**

The model 1 can be written as follows:

$$\Delta r = \begin{cases} & Probability \\ -\gamma & T_{-\gamma}(r) * \Delta t \\ +\gamma & T_{+\gamma}(r) * \Delta t \\ 0 & 1 - \left(T_{+\gamma} + T_{-\gamma}\right) * \Delta t \end{cases} \tag{A1}$$

where $T_{-\gamma}(r) = k2n = k2Nr$ and $T_{+\gamma}(r) = k1*(N-n) = k1\ N\ (1-r)$.

Thus, the evolution of the probability distribution of Model A1 at time $t + \Delta t$, in terms of this distribution at time $t$, can be assessed by Equation A2, known as forward Kolmogorov equation:

$$P(r, t + \Delta t) = P(r - \gamma, t)T_{-\gamma}(r - \gamma)\Delta t + P(r + \gamma, t)T_{+\gamma}(r + \gamma)\Delta t + P(r,t)(1 - T_{-\gamma}(r) - T_{\gamma}(r))\Delta t \tag{A2}$$

In the next step, applying the Taylor series in the first two terms on the right side of the equality A2, around $r$, produces:

$$P(r - \gamma, t)T_{-\gamma}(r - \gamma) = P(r,t)T_{-\gamma}(r) - \gamma \frac{\partial(P(r,t)T_{-\gamma}(r))}{\partial r} + \frac{1}{2}\gamma^2 \frac{\partial 2(P(r,t)T_{-\gamma}(r))}{\partial r^2} + o(\gamma^3) \tag{A3}$$

and

$$P(r + \gamma, t)T_{+\gamma}(r + \gamma) = P(r,t)T_{+\gamma}(r) + \gamma \frac{\partial(P(r,t)T_{+\gamma}(r))}{\partial r} + \frac{1}{2}\gamma^2 \frac{\partial 2(P(r,t)T_{+\gamma}(r))}{\partial r^2} + o(\gamma^3) \tag{A4}.$$

Substituting Equations A3 and A4 into A2, assuming small values for $\gamma$ and $\Delta t$, and simplifying, the expression for $P(r,t)$ can be approximated by the solution of the partial differential equation known as the Fokker-Planck equation, resulting in:

$$\frac{\partial p(r,t)}{\partial t} = -\frac{\partial\left(\gamma\left(T_{+\gamma}(r) - T_{-\gamma}(r)\right)p(r,t)\right)}{\partial r} + \frac{1}{2}\frac{\partial 2\left(\gamma^2\left(T_{+\gamma}(r) + T_{-\gamma}(r)\right)p(r,t)\right)}{\partial r^2} \tag{A5},$$

or even in:

$$\frac{\partial p(r,t)}{\partial t} = -\frac{\partial(\gamma(k1N(1-r) - k2Nr)p(r,t))}{\partial r} + \frac{1}{2}\frac{\partial 2\left(\gamma^2(k1N(1-r) + k2Nr)p(r,t)\right)}{\partial r^2} \tag{A6}.$$

Finally, the stochastic differential equation proposed by Model 2 has the same form as the Fokker-Planck equation (A6), as can be seen in Chapter 4 of Allen (Allen, 2007), for example.

It is important to note that the same stochastic differential equation model (Equation 4) was deduced in a different context by Dangerfield and colleagues (Dangerfield, Kay, & Burrage, 2012). These authors deduced a stochastic differential equation similar to Equation 4 for study of ion channels. In addition, these authors also proposed a reflective edge condition at $r = 0$ and $r = 1$ so that the Equation 4 solution was always in the range [0 1]. Intuitively, the condition of reflective edges means that for $r$ in the range [0 1] the dynamics for this variable follow Equation 4. When $r$ assumes the values 0 or 1, it is rebated to the interval [0 1]. The study by Grasman and van Herwaarden (Grasman & van Herwaarden, 2010) presents a technique for obtaining an approximation of the probability distribution of Equation 4 under the condition of reflective edges used in this article.